\documentclass{moriond}

\def\be{\begin{equation}}
\def\ee{\end{equation}}
\def\bea{\begin{eqnarray}}
\def\eea{\end{eqnarray}}

\def\eg{e.g.~}
\def\fbinv{\,\textrm{fb}^{-1}}

\newcommand{\Photo}{}

\begin{document}
\vspace*{4cm}
\title{Indirect searches at CMS}

\author{Nicholas Smith, on behalf of the CMS Collaboration}

\address{Fermi National Accelerator Laboratory}

\maketitle\abstract{ In these proceedings, we present several new measurements
of Standard Model (SM) processes, in the Higgs sector and beyond, that push the
precision frontier forward at CMS. Results are presented in the context of a
framework parameterizing deviations in Higgs boson couplings, as well as in the
context of SM Effective Field Theory, where new analyses targeting Higgs, top,
and multi-boson processes probe an increasingly diverse set of operators.
Within these frameworks, CMS is efficiently exploring a large space of new
physics models. No significant deviation from SM expectations is found.  }

\section{Introduction}

Indirect searches for new physics are a crucial component of the Compact Muon
Solenoid (CMS)~\cite{CMS:2008xjf} experimental physics program. The basic
premise is that any new resonance or other phenomenon that does not emerge
directly at the energy scale probed the the Large Hadron Collider (LHC) will
nevertheless influence the production rates and kinematic distributions of
Standard Model (SM) particles.  There are several choices to be made in
building a low-energy description of how the hypothetical new physics scenario
would influence these observables.  In this work, we discuss results cast in
two popular approaches: the ``$\kappa$ framework'' and the SM Effective Field
Theory (EFT) framework.  Further details on these approaches appear below.

\section{Higgs measurements}

\subsection{Higgs boson couplings}

Immediately following the Higgs boson
discovery~\cite{ATLAS:2012yve,CMS:2012qbp,CMS:2013btf}, a comprehensive program to
investigate its couplings to other SM particles was laid
out.\cite{LHCHiggsCrossSectionWorkingGroup:2013rie} In time for the 10 year
anniversary of the discovery, CMS has released~\cite{CMS:2022dwd} a combination
of analyses using the full LHC Run-II dataset (2016-2018, up to
$138\,\fbinv$) targeting all major production and decay modes, that
achieves an overall Higgs boson signal strength $\mu =
1.002\pm0.057$ with respect to the SM expectation. 

Given this precision, it is clear that there is a wealth of information to be
gained by looking differentially in the production and decay modes for
deviations from the SM.  In this result, the relative signal strengths of each
production and decay mode are reported. However, such a parameterization does
not capture the correlated shifts in production cross sections and decay
branching fractions that would be expected from deviations in the coupling
between the Higgs boson and other SM particles. The $\kappa$ framework allows
each interaction vertex to be modified by a free parameter, nominally 1 (\eg
$\kappa_t$ scales the ttH interaction vertex), which consistently scales
deviations from SM expectations in both Higgs boson production and decay. In
this result, limits are placed simultaneously on
$\kappa_W,\kappa_Z,\kappa_t,\kappa_b,\kappa_\tau,\kappa_\mu,\kappa_\gamma,\kappa_g,$
and $\kappa_{Z\gamma}$ (the latter 3 couplings are effective couplings,
integrating out a loop process.) Many coupling modifiers are constrained to be
within 10\% of unity. Projections of expected constraints on these couplings at
the High-Luminosity LHC are also presented, showing an ultimate precision of a
few percent---as low as 1\% with improvements in systematic uncertainties such
as those from proton PDFs---is achievable.

\subsection{Rare Higgs boson couplings}

Under the SM hypothesis, the $H\bar{f}f$ coupling is proportional to the
fermion mass, hence observing its coupling to lighter fermions is challenging.
For charm quarks, CMS has excluded production rates above 14 times the SM cross
section at 95\% confidence level (CL) for the $VH(c\bar{c})$
process.\cite{CMS:2022psv}  An excess of events leads to a two-sided observed
95\% CL interval on the coupling modifier, $1.1 \leq \kappa_c \leq 5.5$.

There is little chance for direct observation of Higgs boson decays to  $u,d,$
or $s$ quarks at SM rates. Upper limits on the coupling can be placed through
indirect methods, though they require additional theoretical assumptions.
Constraints on the primary Higgs boson couplings~\cite{CMS:2022dwd} do not imply any
constraint on light quark $\kappa$ values due to a degeneracy: one can preserve the
signal strength for the primary production-decay processes by scaling all other
Higgs boson couplings in response to scaling the light quark
coupling.\cite{Coyle:2019hvs} The degeneracy can be broken if the Higgs boson total
width ($\Gamma_H$) is known, or some other constraint is placed on the total
width, such as restricting $\kappa_W,\kappa_Z\leq 1$.  CMS has indirectly
measured $\Gamma_H$,\cite{CMS:2022ley} under additional theoretical
assumptions.

Direct limits on light quark couplings can in principle be placed via searches
for rare Higgs boson decays to $Z\rho$ and $Z\phi$,\cite{CMS:2020ggo} though the
leading SM contribution to such processes do not involve a $Hq\bar{q}$ vertex.
Once can consider such vertices on the production side by searching for the
$q\bar{q}\rightarrow H+\gamma$ process.\cite{Aguilar-Saavedra:2020rgo} By
tagging $H(WW)$ decays, this process is used by CMS to place upper limits on
$q\bar{q} \rightarrow H+\gamma \rightarrow e\mu\gamma$ production cross
sections and associated $\kappa_q$ values for $q=u,d,s$, and $c$
quarks.\cite{CMS:2023jpy} The limits are placed under the assumption that
$\Gamma_H$ increases in the required way to preserve agreement with data in the
primary Higgs boson production channels.

\subsection{Higgs boson self-coupling}
\label{sec:hh}

Beyond-SM (BSM) modifications to the Higgs boson self-coupling can similarly be expressed as a
generic scaling of the trilinear vertex, $\kappa_\lambda$. CMS has used a
combination of several channels sensitive to di-Higgs production to constrain
$-1.24 < \kappa_\lambda < 6.49$.\cite{CMS:2022dwd} More recent di-Higgs
measurements that have not yet entered the combination include the
$H(WW)H(\gamma\gamma)$~\cite{CMS:2022rgm} and $H(bb)H(WW)$~\cite{CMS:2023qiw}
decay topologies for gluon-initiated production, as well as the $VHH(4b)$
associated-production~\cite{CMS:2023ehl} topology.

\begin{figure}[b]
  \centering
  \includegraphics[height=4.5cm,keepaspectratio]{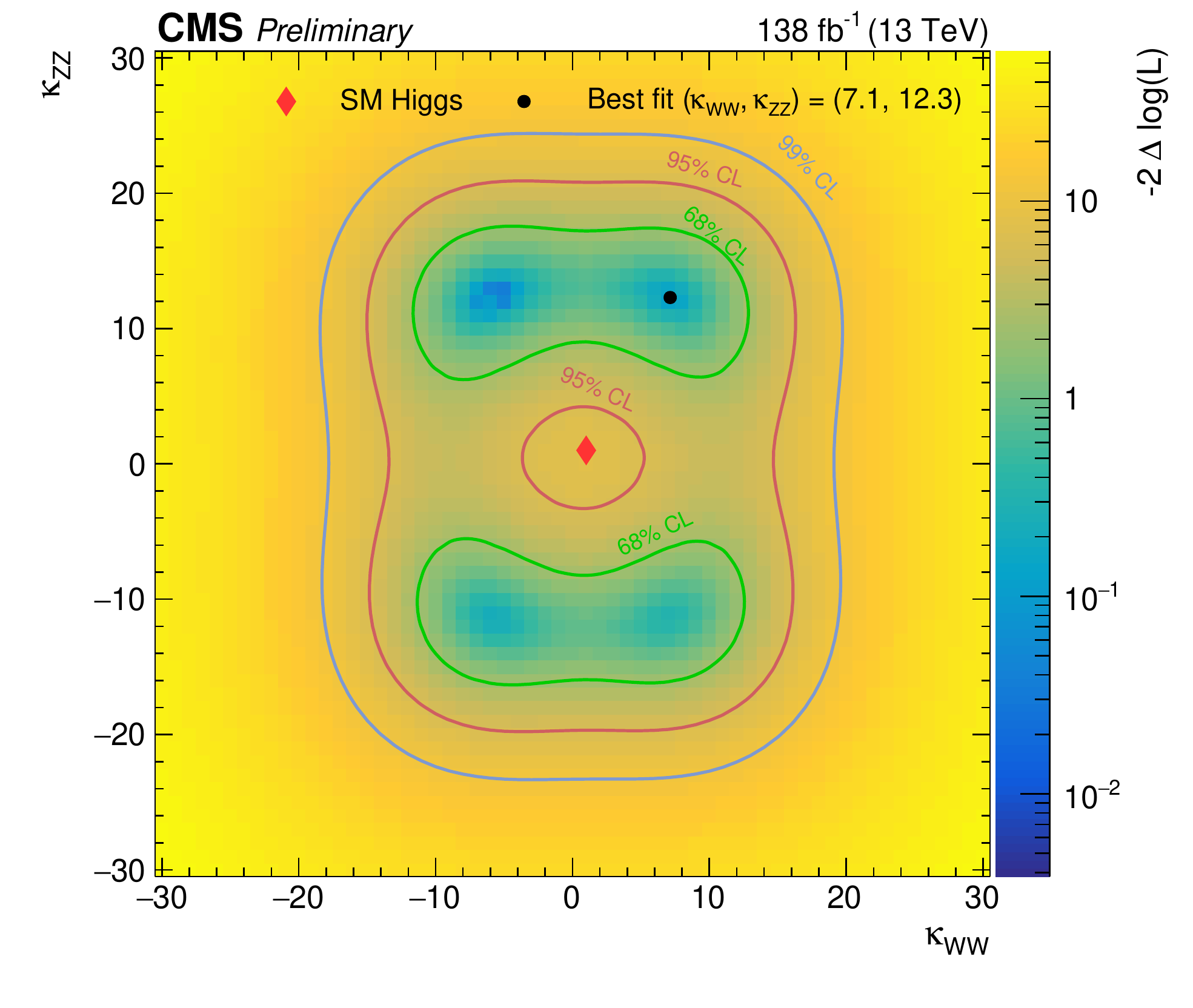}
  \includegraphics[height=4.5cm,keepaspectratio]{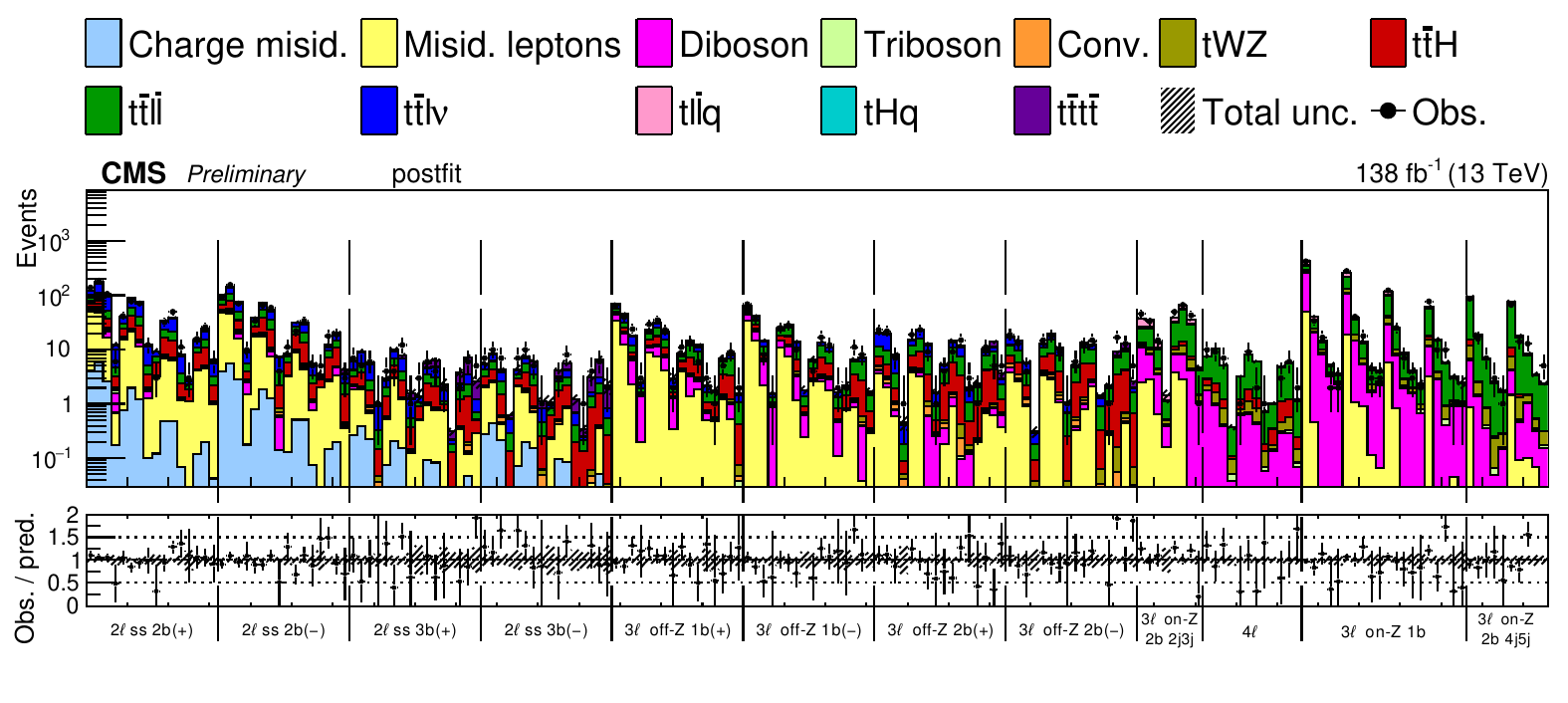}
  \caption{
    Left: Observed likelihood scans of $\kappa_{WW}$ versus $\kappa_{ZZ}$, with other SM couplings fixed to the SM predicted strength, as determined in the $VHH(4b)$ analysis discussed in Sec.~\ref{sec:hh}.
    Right: Expected and observed yields for all regions considered in the top multi-lepton analysis discussed in Sec.~\ref{sec:topml}, demonstrating the large number of observables needed to constrain 26 Wilson coefficients simultaneously.
  }
  \label{fig:vhh}
\end{figure}

Though these channels are not among the most sensitive to deviations
in $\kappa_\lambda$, they are nevertheless very effective probes of several
other anomalous couplings, such as BSM point-like interactions between two
Higgs bosons and two gluons, or deviation from the SM expectation in the $HHVV$
interaction.  Uniquely in the case of the $VHH(4b)$
analysis, potential deviations in the $HHWW$ and $HHZZ$
couplings are parameterized and constrained independently, as shown in Fig.~\ref{fig:vhh} left.

\section{SM Effective Field Theory}

From its inception, the $\kappa$ framework was seen as an interim approach to
parameterizing deviations from SM Higgs boson behavior, due to deficiencies such as
the lack of a consistent treatment of higher-order contributions, a coupling
tensor structure restricted to that of the SM, and no mechanism for running of
couplings with renormalization scale. In addition, this framework is of no use
beyond Higgs boson characterization. In contrast, EFT techniques can allow a
consistent treatment of all the above considerations. In particular, the SM
EFT~\cite{Grzadkowski:2010es,Degrande:2012wf,Brivio:2017vri} enumerates BSM
couplings associated with all non-renormalizable operators involving SM fields
up to a given mass dimension which respect the SM gauge symmetries, as the
low-energy limit of an unspecified BSM model at scale $\Lambda \gg v \approx 246$GeV. In
this section, we highlight several recent CMS results that either directly set
unique constraints on SMEFT parameters or could easily be re-interpreted to do
so. We group results by the dimension-6 operators they may probe.

\subsection{Dimension-6 $\bar{f}fXY$ operators}

A full Run-II analysis searching for lepton-flavor violating (LFV) Higgs boson decays
to the $e\mu$ final state has set an observed upper limit of
$\mathcal{B}(H\rightarrow e\mu) < 0.44\times 10^{-4}$ at 95\%
CL.\cite{CMS:2023pqk} One can interpret this result as a limit on the
coefficient of operators such as $(\phi^\dagger \phi) (\bar{l}_1 e_2 \phi)$
(along with permutations of the flavor indices.)~\cite{Harnik:2012pb} In
general, direct searches for LFV Higgs boson decays are less powerful than indirect
constraints, \eg via $\mu\rightarrow e\gamma$ branching ratio measurements.
Yet, in the case of an excess in the latter, these direct constraints would be
critical in determining if the Higgs boson was responsible.

A recent CMS result probes flavor-changing neutral currents in the quark
sector involving anomalous $tc\gamma$ and $tu\gamma$ interactions via the
$e/\mu +\gamma+$jets final state, setting limits on the branching ratio
$t\rightarrow u/c + \gamma$ of order $10^{-5}$.\cite{CMS:2023tir} Results are
interpreted in a top anomalous couplings
scenario,\cite{Aguilar-Saavedra:2008nuh} that can be re-cast in the Warsaw
basis~\cite{Grzadkowski:2010es} as limits on the coefficients of operators such
as $(\bar{q}_3\sigma^{\mu\nu}u_2) \sigma^i \tilde{\phi}W_{\mu\nu}^i$ and
$(\bar{q}_3\sigma^{\mu\nu}u_2) \tilde{\phi}B_{\mu\nu}$ (along with flavor index
permutations.) A key feature exemplified by this analysis is that all
significant backgrounds are determined via a data-driven estimate, which is
important to insulate the extraction of limits on anomalous couplings in the
targeted topology from possible BSM deviations in the backgrounds.

\subsection{Dimension-6 four-fermion operators}

Charged LFV is probed in the top sector in a recent CMS result, setting limits
on dimension-6 four-fermion operators contributing to anomalous $e\mu t (u/c)$
contact interactions. Grouping operators with a common Lorentz structure
together, limits are placed on 6 coefficients from as low as $0.02\,
\textrm{TeV}^{-2}$ for $e\mu tu$ tensor coupling, to $0.3\,\textrm{TeV}^{-2}$ for the $e\mu tc$
scalar contact interaction.\cite{CMS:2023phe} As there is no SM interference, a
potential deviation appears only at order $\Lambda^{-4}$ in the new physics
scale.  The interactions are probed in final states targeting a di-top system
where one top decays $t \rightarrow e\mu (u/c)$ and single-top production in
association with an electron and muon. The latter channel is found to drive the
sensitivity.

Limits on four-lepton contact operators can be probed effectively via searches
for rare $Z\rightarrow 4\ell$ decays.\cite{Boughezal:2020klp} A recent CMS
result uses a shape analysis of the $4\mu$ invariant mass near the $Z$ pole to
set an upper limit on the ratio of branching fractions
$\mathcal{B}(Z\rightarrow \tau\tau\mu\mu) / \mathcal{B}(Z\rightarrow 4\mu) <
6.2$ at 95\% CL.\cite{CMS:2023nkg}

CMS has observed four-top production,\cite{CMS:2023ica} which provides yet
another useful handle to constrain the space of possible four-fermion contact
interactions. Further details on the significant analysis improvements made to
reach this milestone were presented in another talk.

\subsection{Top multi-lepton}
\label{sec:topml}

A recent CMS result uses a finely-categorized analysis of final states
involving top quark decays with additional leptons to set non-degenerate
constraints on 26 dimension-6 Wilson coefficients
simultaneously.\cite{CMS:2023ixc} The analysis proceeds via progressively
categorizing: lepton multiplicity ($\ell^\pm \ell^\pm$, $\ell^\pm\ell^\mp
\ell$, and $4\ell$); on- or off-shell $Z$ for three-lepton events; b-tagged jet
multiplicity; charge sum; jet multplicity; and finally a category-dependent
kinematic variable chosen to optimize sensitivity to the leading operator
coefficient within that category. This fine categorization, shown in
Fig.~\ref{fig:vhh} right, enables the analysis to effectively constrain the
high-dimensional model space. All significant single- and di-top processes
that include additional leptons are simulated at leading-order in QCD, and the
effects of all considered EFT operator coefficients are parameterized at the
per-bin level in the likelihood model used to extract the limits.

\section{Conclusions}

CMS is continuing to produce many important results probing a wide range of new
physics model space via indirect methods. So far, no significant deviation from
the SM expectation is found. However, with a common language, combinations of a
diverse set of analyses may be able to uncover subtle signs of new physics.
Enthusiasm for the dimension-6 SMEFT parameterization is growing, and the
technology and tooling is now available. As such, this is a promising direction
to pursue for indirect searches at the LHC in the near future.

\section*{Acknowledgments}

The author would like to thank the CMS Collaboration and the organizers of the
57\textsuperscript{th} Recontres de Moriond for the opportunity to present
these results.  This work was partially supported by Fermilab operated by Fermi
Research Alliance, LLC under Contract No. DE-AC02-07CH11359 with the United
States Department of Energy, and by the National Science Foundation under
Cooperative Agreement PHY-2121686.

\end{document}